\documentclass[aps,prl,reprint,amsmath,amssymb,nofootinbib,superscriptaddress]{revtex4-2}
\usepackage{graphicx}
\usepackage{bm}
\usepackage{xcolor}
\usepackage[colorlinks=true,linkcolor=blue!60!black,citecolor=blue!60!black,urlcolor=blue!60!black]{hyperref}
\usepackage{orcidlink}

\newcommand{\dd}{\mathrm{d}}
\newcommand{\Te}{T_{\mathrm{e}}}
\newcommand{\nion}{n_{\mathrm{i}}}
\newcommand{\nc}{n_{\mathrm{c}}}
\newcommand{\nic}{n_{\mathrm{i,c}}}
\newcommand{\Zb}{\langle Z\rangle}
\newcommand{\lamL}{\lambda_{\mathrm{L}}}
\newcommand{\tL}{t_{\mathrm{L}}}
\newcommand{\cs}{c_{\mathrm{s}}}
\newcommand{\kabs}{\bar{\kappa}}
\newcommand{\ebar}{\bar{\varepsilon}}
\newcommand{\kDk}{\kappa_{D,k}}
\newcommand{\relL}{\varepsilon_{L2}}
\newcommand{\etot}{\varepsilon_{\mathrm{tot}}}
\newcommand{\eib}{\varepsilon_{\mathrm{ib}}}
\newcommand{\um}{\,\mu\mathrm{m}}

\begin{document}

\title{Data-driven non-contiguous photon-energy groups for fast multigroup radiation transport in line-dominated plasmas}

\author{Shinsuke Fujioka\,\orcidlink{0000-0001-8406-1772}}
\email{fujioka.shinsuke.ile@osaka-u.ac.jp}
\affiliation{Institute of Laser Engineering, The University of Osaka, 2-6 Yamadaoka, Suita, Osaka 565-0871, Japan}
\author{Yuji Takagi\,\orcidlink{0000-0002-0922-2318}}
\affiliation{Institute of Laser Engineering, The University of Osaka, 2-6 Yamadaoka, Suita, Osaka 565-0871, Japan}
\author{Nozomi Tanaka\,\orcidlink{0000-0002-8392-2224}}
\affiliation{Institute of Laser Engineering, The University of Osaka, 2-6 Yamadaoka, Suita, Osaka 565-0871, Japan}
\affiliation{National Institute for Fusion Science, 322-6 Oroshi, Toki, Gifu 509-5292, Japan}
\author{Akira Sasaki\,\orcidlink{0000-0002-5094-7361}}
\affiliation{Institute of Laser Engineering, The University of Osaka, 2-6 Yamadaoka, Suita, Osaka 565-0871, Japan}
\author{Yu Yamamoto\,\orcidlink{0000-0002-1549-6382}}
\affiliation{Institute of Laser Engineering, The University of Osaka, 2-6 Yamadaoka, Suita, Osaka 565-0871, Japan}
\author{Muhammad Fauzan Syahbana}
\affiliation{Institute of Laser Engineering, The University of Osaka, 2-6 Yamadaoka, Suita, Osaka 565-0871, Japan}
\author{Katsunobu Nishihara\,\orcidlink{0000-0002-5572-1169}}
\affiliation{Institute of Laser Engineering, The University of Osaka, 2-6 Yamadaoka, Suita, Osaka 565-0871, Japan}
\affiliation{Faculty of Engineering, Osaka Metropolitan University, 3-3-138 Sugimoto, Sumiyoshi-ku, Osaka 558-8585, Japan}
\author{Tomoyuki Johzaki\,\orcidlink{0000-0002-5738-4661}}
\affiliation{Graduate School of Advanced Science and Engineering, Hiroshima University, 1-4-1 Kagamiyama, Higashihiroshima, Hiroshima 739-8527, Japan}
\affiliation{Institute of Laser Engineering, The University of Osaka, 2-6 Yamadaoka, Suita, Osaka 565-0871, Japan}
\author{Atsushi Sunahara\,\orcidlink{0000-0001-7543-5226}}
\affiliation{Center for Materials Under eXtreme Environment (CMUXE), School of Nuclear Engineering, Purdue University, 500 Central Drive, West Lafayette, Indiana 47907, USA}
\affiliation{Institute of Laser Engineering, The University of Osaka, 2-6 Yamadaoka, Suita, Osaka 565-0871, Japan}
\author{Masashi Yoshimura\,\orcidlink{0000-0002-0770-1251}}
\affiliation{Institute of Laser Engineering, The University of Osaka, 2-6 Yamadaoka, Suita, Osaka 565-0871, Japan}

\date{\today}

\begin{abstract}
%
Radiation transport in line-dominated plasmas, such as extreme ultraviolet (EUV) light source plasma, laser-fusion hohlraums, and stellar interiors, needs too many photon-energy groups for accurate radiation-hydrodynamic simulations.
Our data-driven method assigns the bins of an opacity table to $K$ non-contiguous groups with fixed spectral shapes that never couple.
Existing multigroup codes can use it by replacing the opacity table.
In a radiation-hydrodynamic code, 30 such groups replaced 151 conventional groups; the energy conversion efficiency from laser to EUV changed by less than 0.5\% in absolute value.
\end{abstract}

\maketitle

Radiation carries energy in hot dense matter.\cite{Mihalas1984} Examples are laser-fusion hohlraums,\cite{Lindl1995,Lindl2004,AbuShawareb2024} stellar interiors and envelopes,\cite{Rogers1994,Seaton1994,Iglesias1996,Bailey2015} and the tin plasmas used as the 13.5-nm light source for extreme-ultraviolet (EUV) lithography.\cite{Nishihara2008,OSullivan2015,Versolato2019}
In these plasmas, mid- and high-$Z$ ions have thousands of overlapping bound--bound transitions.
The opacity changes by orders of magnitude within less than 1~eV, between optically thick line cores and optically thin windows.\cite{Bauche1979}
Radiation either escapes through the windows or is reabsorbed in the lines. This determines the emitted spectrum and finally the hydrodynamics.\cite{Fujioka2005}
In Sn, the spectrum is dominated by unresolved transition arrays (UTAs) of the $4d$--$4f$ and $4p$--$4d$ transitions of Sn$^{8+}$--Sn$^{14+}$.\cite{Sasaki2026}
A multigroup diffusion calculation needs about $10^{3}$ photon-energy groups to resolve them; the table used here has $G=1500$ fine bins between 0 and 1500~eV.
The cost of multigroup diffusion is proportional to the number of groups.
Multidimensional codes therefore use only a few tens of \emph{contiguous} bands, with Planck-mean and Rosseland-mean opacities in each band.\cite{Mihalas1984}
This approximation fails when a band contains both thick lines and thin windows.
Groups that are not contiguous in photon energy are an old remedy.
Opacity binning puts frequencies with a similar \emph{value} of the opacity into one group.\cite{Nordlund1982,Mihalas1984,Auer2000,Wray2006}
Correlated-$k$ methods reorder the spectrum by the absorption coefficient,\cite{Goody1989,Modest2002} and their multigroup form sorts the frequencies by the \emph{state dependence} of the opacity.\cite{Zhang2003}
Non-contiguous regrouping also speeds up Monte Carlo transport.\cite{Wollaeger2014}
In all these methods, the groups are built from the opacity alone, and the spectrum is flat inside a group.
In a self-absorbed line spectrum that is not in local thermodynamic equilibrium, however, the quantity to be represented is the radiation field. Its structure inside a group is set by transport as well as by the local opacity.
In fluid dynamics, similar problems have been solved by data-driven model reduction, such as proper orthogonal decomposition (POD)\cite{Berkooz1993} and hyper-reduction.\cite{Chaturantabut2010}

In this Letter, we apply such a reduction to multigroup radiation transport.
We require that the reduced model is as cheap as a multigroup code.
This requirement leads to a simple result: $K$ \emph{generalized groups}.
Each generalized group is a set of fine bins that do not need to be neighbors in photon energy, and it has a fixed spectral shape.
The training data show which parts of the spectrum change together when the temperature, the density, and the depth change. Only bins that change together are put into one group.
Compared with the opacity-based methods above, the groups are built from the radiation field, and the shape inside each group is taken from the data.
The reduced model has the same structure as a usual multigroup flux-limited diffusion (FLD) code with $K$ uncoupled groups.
It can therefore replace the opacity table of an existing code.

\textit{Reference problem.}---We consider a one-dimensional, planar, steady, isothermal Sn plasma that expands from the critical surface of a laser with wavelength $\lamL$ (in $\mu$m).
The critical density is $\nc=1.1\times10^{21}/\lamL^{2}$~cm$^{-3}$.
The ion density at the critical surface, $\nic$, is given by $\Zb(\Te,\nic)\,\nic=\nc$, where the mean charge $\Zb(\Te,\nion)$ is taken from the table.
The density profile is $\nion(x)=\nic\exp(-x/L)$ for $0\le x\le x_{\mathrm b}=500\um$, with $L=\cs\tL$ and $\cs=\sqrt{\Zb_{\mathrm c}\Te/m_{\mathrm{Sn}}}$.
The time $\tL$, of the order of the laser pulse duration, only sets the scale length.
We write $E\equiv h\nu$ for the photon energy and $\Delta E_g$ for the width of fine bin $g$.
The radiation energy density $U_g$ in bin $g$ obeys
\begin{align}
  -\frac{\dd}{\dd x}\!\left[D_g\frac{\dd U_g}{\dd x}\right]+c\kappa_g U_g&=\varepsilon_g\Delta E_g,
  \label{eq:mg}\\
  D_g=\frac{c\,\lambda(R_g)}{\kappa_g},\qquad R_g&=\frac{|\dd U_g/\dd x|}{\kappa_g U_g}.\nonumber
\end{align}
The opacity $\kappa_g$ (cm$^{-1}$) and the emissivity $\varepsilon_g$ (erg\,cm$^{-3}$\,s$^{-1}$\,eV$^{-1}$) are interpolated in $(\log\Te,\log\nion)$.
We use the Larsen flux limiter $\lambda(R)=(3^{n}+R^{n})^{-1/n}$ with $n=2$\cite{Olson2000} and Marshak vacuum boundaries.
The plasma is isothermal and is not in local thermodynamic equilibrium at these densities ($\varepsilon_g\neq4\pi\kappa_gB_g$).
We solve Eq.~\eqref{eq:mg} with finite volumes ($\Delta x=1\um$) and Picard iteration for the flux limiter (Supplemental Material).
The output is the spectral flux $F_E$ that leaves the plasma at $x_{\mathrm b}$. Its value in bin $g$ is $F_g$.
The result of the 1500-group calculation, $F_g^{\mathrm{ref}}$, is the reference for all reduced models.
Below we use the spectral energy density $u_g=U_g/\Delta E_g$ and the inner product $\langle u,v\rangle=\sum_g u_gv_g\Delta E_g$.

\textit{Mode expansion is accurate but slow.}---A standard data-driven reduction writes the spectrum at each position as a sum of a few fixed spectral shapes $\phi_k(g)$ ($k=1,\dots,K$) with amplitudes that depend on position, $u_g(x)\simeq\sum_k a_k(x)\phi_k(g)$.
The residual of Eq.~\eqref{eq:mg} is then made orthogonal to every shape (Galerkin projection). This gives
\begin{equation}
  -\frac{\dd}{\dd x}\!\left[\mathfrak{D}\frac{\dd\bm a}{\dd x}\right]+c\,\mathcal{K}\bm a=\bm{\mathcal S},\quad
  \mathcal{K}_{kl}=\sum_g\phi_k\phi_l\kappa_g\Delta E_g,
  \label{eq:galerkin}
\end{equation}
with $\mathcal S_k=\sum_g\phi_k\varepsilon_g\Delta E_g$ and $\mathfrak D_{kl}=\sum_g\phi_k\phi_lD_g\Delta E_g$.
If the shapes are the POD modes of the training spectra, only a few modes are needed. Forty-one modes contain 99.9\% of the ($\Delta E$-weighted) energy, and $K=100$ reproduces the reference boundary spectrum with a median relative $L^{2}$ error of 1\%.
However, this model is slow.
Each POD mode extends over the whole spectrum, so absorption at any photon energy couples all amplitudes.
The matrices $\mathcal K$ and $\mathfrak D$ are dense, and Eq.~\eqref{eq:galerkin} is a $K\times K$ block system.
Its cost is proportional to $K^{2}$--$K^{3}$, which is more than the cost of $G$ independent scalar equations when $K\gtrsim\sqrt G\approx40$.
In addition, the flux limiter must be calculated for all $G$ bins at every nonlinear iteration.
In our tests, the POD--Galerkin model with $K=100$ was more than 10 times \emph{slower} than the reference, even with tabulated operators and an approximated limiter (Table~\ref{tab:compare}; Supplemental Material).
The problem is the coupling between the modes, not the number of modes.

\textit{Uncoupled modes are generalized groups.}---We therefore look for modes that never couple. This means that $\mathcal K$ must be diagonal for \emph{every} plasma state.
The opacity $\kappa_g(\Te,\nion)$ changes with the state in a different way in each bin.
The condition can therefore be satisfied only if the supports $S_k=\{g:\phi_k(g)\neq0\}$ do not overlap:
\begin{equation}
  \phi_k(g)=\begin{cases}w_k(g), & g\in S_k,\\ 0, & g\notin S_k,\end{cases}\quad
  \sum_{g\in S_k}\!w_k^{2}\Delta E_g=1,\ w_k\ge0.
  \label{eq:disjoint}
\end{equation}
Each fine bin then belongs to exactly one mode.
We call $S_k$ a \emph{generalized group} and $w_k$ its \emph{intra-group shape}.
There are only two differences from a usual multigroup model: $S_k$ does not need to be contiguous in photon energy, and $w_k$ does not need to be flat.
A thick line core and the thin window next to it can be put into different groups, and distant lines with a similar state dependence can be put into the same group.
With Eq.~\eqref{eq:disjoint}, Eq.~\eqref{eq:galerkin} becomes $K$ independent scalar equations,
\begin{align}
  -\frac{\dd}{\dd x}\!\left[\bar D_k\frac{\dd a_k}{\dd x}\right]+c\kabs_k a_k&=\ebar_k,
  \label{eq:fewgroup}\\
  \kabs_k=\!\sum_{g\in S_k}\!w_k^{2}\kappa_g\Delta E_g,\quad
  \ebar_k&=\!\sum_{g\in S_k}\!w_k\varepsilon_g\Delta E_g,\nonumber\\
  \frac{1}{\kDk}&=\!\sum_{g\in S_k}\!\frac{w_k^{2}\Delta E_g}{\kappa_g}.\nonumber
\end{align}
Here $\kabs_k$ is a Planck-type mean and $\kDk$ is a Rosseland-type (harmonic) mean, both weighted by the shape. They become the usual band means when $w_k$ is flat.
The flux limiter and the Marshak conditions are applied to each group, with $\bar D_k=c\lambda(R_k)/\kDk$ and $R_k=|\dd a_k/\dd x|/(\kDk a_k)$.
The three quantities $\kabs_k$, $\ebar_k$, and $\kDk$ are calculated once on the $(\Te,\nion)$ grid of the original table. The new table is smaller by a factor $K/G$ (1/15 for $K=100$).
The emitted spectrum is reconstructed as $F_g=\bar g_k a_k(x_{\mathrm b})w_k(g)$ for $g\in S_k$, where $\bar g_k$ is the Marshak boundary coefficient.

\textit{Which bins change together.}---We now have to choose $S_k$ and $w_k$.
The training data are spectral energy densities $u^{(s)}$, normalized to $\int u\,\dd E=1$. They are sampled every $5\um$ inside the plasma described above, for 5\% of the parameter grid, and are calculated with the reference code.
For each fine bin $g$, the vector of its values $\sqrt{\Delta E_g}\,u_g^{(s)}$ over all samples $s$ shows how the bin becomes brighter or darker when the plasma state and the depth change.
The Gram matrix
\begin{equation}
  \mathsf A=\mathsf W^{1/2}\Big(\frac1n\sum_s u^{(s)}u^{(s)\mathsf T}\Big)\mathsf W^{1/2},\qquad \mathsf W=\mathrm{diag}(\Delta E_g),
  \label{eq:gram}
\end{equation}
contains the inner products of these vectors.
The diagonal element $\mathsf A_{gg}$ is the energy of bin $g$ in the training set.
The cosine $C_{gg'}=\mathsf A_{gg'}/\sqrt{\mathsf A_{gg}\mathsf A_{g'g'}}$ is 1 when bins $g$ and $g'$ always change in proportion, and 0 when they change independently.
A set of bins that always change in proportion can be described by one amplitude and one fixed shape. Such a set is a generalized group.
In this case, the block $\mathsf A_{S_kS_k}$ has rank one, and its eigenvector is the shape.
The reconstruction error of a basis with non-overlapping supports is $\sum_k[\mathrm{tr}\,\mathsf A_{S_kS_k}-\lambda_{\max}(\mathsf A_{S_kS_k})]$.
The best partition therefore makes every block as close to rank one as possible. Only the $G\times G$ matrix $\mathsf A$ is needed, not the samples.
We find an approximate solution in two steps.
First, the bins are clustered by weighted kernel $k$-means\cite{Dhillon2004,Arthur2007} with kernel $C_{gg'}$ and weight $\mathsf A_{gg}$.
Second, the partition is refined. The largest eigenvector $v_k$ of each block gives the shape $w_k=v_k/\sqrt{\Delta E_g}$; it is non-negative by the Perron--Frobenius theorem. Each bin is then moved to the group whose shape explains its energy best, $\arg\max_k(\mathsf A_{g,S_k}v_k)^{2}/\lambda_{\max,k}$. These two operations are repeated until the partition does not change (fewer than ten times).
We measure the quality of a partition by the \emph{explained energy fraction} $\eta=\sum_k\lambda_{\max,k}/\mathrm{tr}\,\mathsf A$; $1-\eta$ is the relative mean-square reconstruction error.
Because $w_k\ge0$, the reconstructed spectra are never negative.

\begin{figure}[tb]
  \centering
  \includegraphics[width=\linewidth]{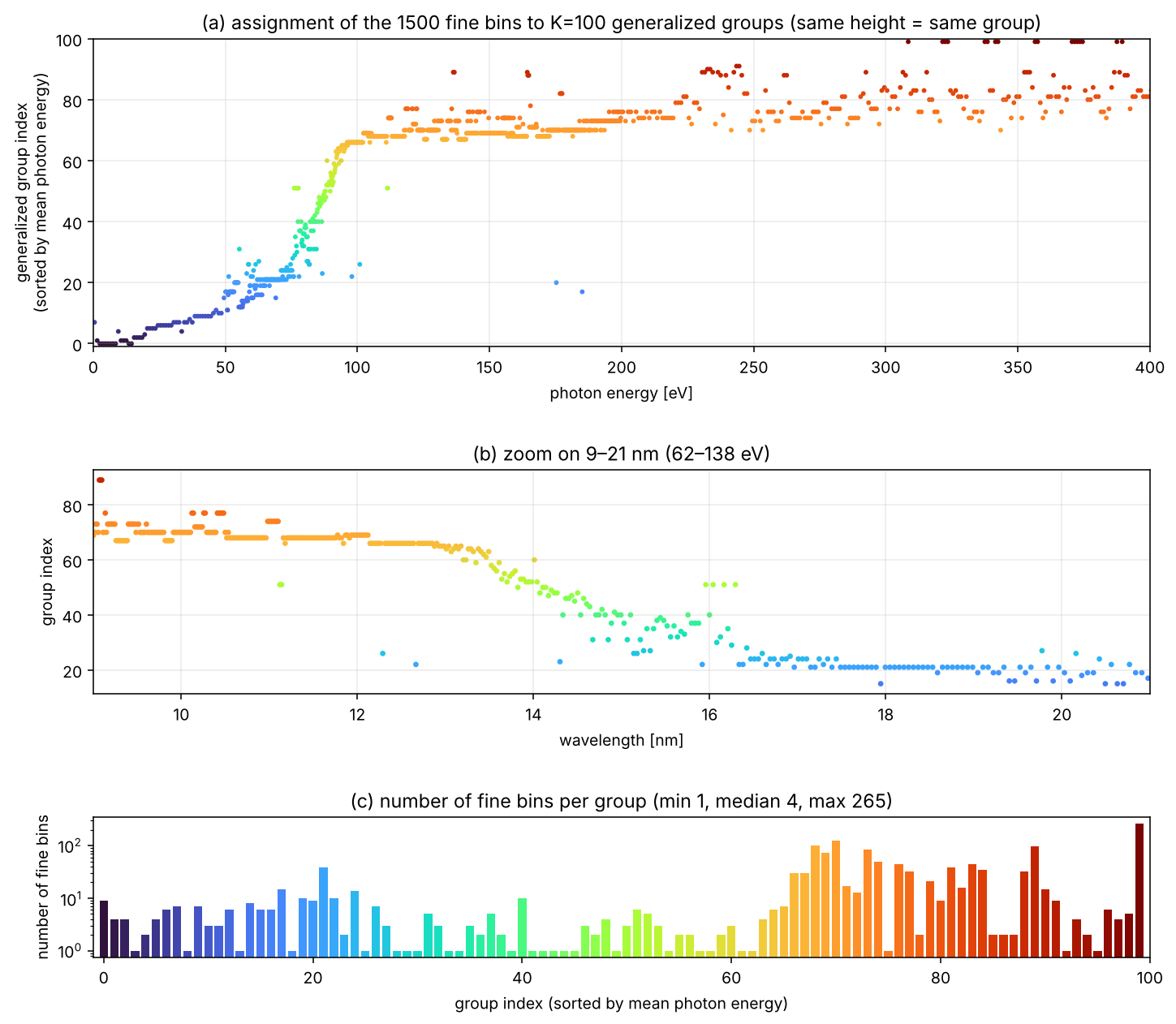}
  \caption{Assignment of the 1500 fine bins to $K=100$ generalized groups. (a)~Group index of each bin as a function of its photon energy. The groups are ordered by their mean photon energy. Points at the same height belong to the same group. (b)~Enlarged view of 9--21~nm. (c)~Number of fine bins in each group.}
  \label{fig:partition}
\end{figure}

\textit{Numerical experiment.}---The parameter grid is $\Te=5$--250~eV (step 5~eV), $\lamL=1$--11$\um$ (step 1$\um$), and $\tL=1$--100~ns (step 5~ns). This gives 11\,550 cases.
We used 577 cases (5.0\%, 57\,700 spectra) for training, 12 for validation (to choose $K$), and 20 for testing.
We evaluate three errors of the boundary spectra of the test cases.
The first is the weighted relative $L^{2}$ error, $\relL=[\sum_g(F_g-F_g^{\mathrm{ref}})^{2}\Delta E_g/\sum_g(F_g^{\mathrm{ref}})^{2}\Delta E_g]^{1/2}$, which is sensitive to line shapes.
The second is the relative error of the total flux, $\etot$.
The third is the relative error of the in-band flux within 13.5~nm~$\pm1\%$, $\eib$.
For comparison, we ran the same solver with $K$ contiguous bands of equal training energy, (a)~with the same rank-one shapes and (b)~with flat shapes. Model (b) corresponds to a conventional multigroup code.
We also built non-contiguous groups from the opacity table alone: (c)~from the value of $\kappa_g$ at one state (opacity binning), and (d)~from the state dependence and the value of $\kappa_g$ over the whole table (multigroup $k$-distribution type). The definitions are given in the Supplemental Material.

\textit{Structure of the groups.}---Figure~\ref{fig:partition} shows the partition for $K=100$.
Below about 50~eV, the groups are almost contiguous bands.
In the $4d$--$4f$ UTA region of 60--130~eV (10--20~nm), the groups are very narrow, and some contain only one bin. Line cores and the windows between them are in different groups.
Fifty-five of the 100 groups are in the 10--20~nm range, which contains 62\% of the training energy.
Above about 150~eV, bins that are far apart in photon energy are put into the same group. The group size is between 1 and 265 bins (median 4).

The errors decrease approximately as $1/K$ (Supplemental Material).
The explained energy fraction is $\eta=98.5\%$ for $K=100$, lower than that of POD with the same $K$ (99.997\%). This is the drawback of having no coupling.
The flux limiter calculated for each group gives the same errors as the limiter calculated for each fine bin, so no fine-bin information is needed during the calculation.

\begin{table*}[tb]
  \caption{Accuracy and computing time for the 20 test cases (one core, Python/NumPy). ``med'' is the median and ``p90'' is the 90th percentile; $\eta$ is the explained energy fraction. The reference time includes 0.7~s for the interpolation of the 1500-bin table; the solver parts alone differ by a factor of about $15\approx G/K$.}
  \label{tab:compare}
  \begin{ruledtabular}
  \begin{tabular}{lcccccc}
    Method & $\eta$ & $\relL$ med & $\etot$ med & $\etot$ p90 & $\eib$ p90 & Time (s) \\
    \hline
    1500-group reference & -- & -- & -- & -- & -- & 1.03 \\
    POD--Galerkin, 100 coupled modes & 0.99997 & 0.036 & 0.006 & 0.014 & 0.048 & 12 \\
    100 groups $\times$ 3 coupled shapes & 0.999 & 0.038 & 0.002 & 0.015 & 0.018 & 0.75 \\
    \hline
    \textbf{This work, 100 non-contiguous groups} & 0.985 & 0.149 & 0.021 & 0.080 & 0.071 & 0.018 \\
    100 contiguous bands, same rank-one shapes & 0.934 & 0.314 & 0.194 & 0.446 & 0.070 & 0.017 \\
    100 contiguous bands, flat shapes (conventional) & 0.649 & 0.552 & 0.137 & 0.864 & 0.045 & 0.016 \\
    100 groups from $\kappa$ value (opacity binning), flat shapes & 0.410 & 0.769 & 0.255 & 0.822 & 2.11 & 0.020 \\
    100 groups from $\kappa$ state dependence, flat shapes & 0.897 & 0.367 & 0.033 & 0.111 & 0.199 & 0.017 \\
    100 groups from $\kappa$ state dependence, rank-one shapes & 0.972 & 0.212 & 0.023 & 0.086 & 0.197 & 0.017 \\
    This work, 400 non-contiguous groups & 0.999 & 0.041 & 0.004 & 0.038 & 0.007 & 0.051 \\
  \end{tabular}
  \end{ruledtabular}
\end{table*}

\begin{figure}[tb]
  \centering
  \includegraphics[width=\linewidth]{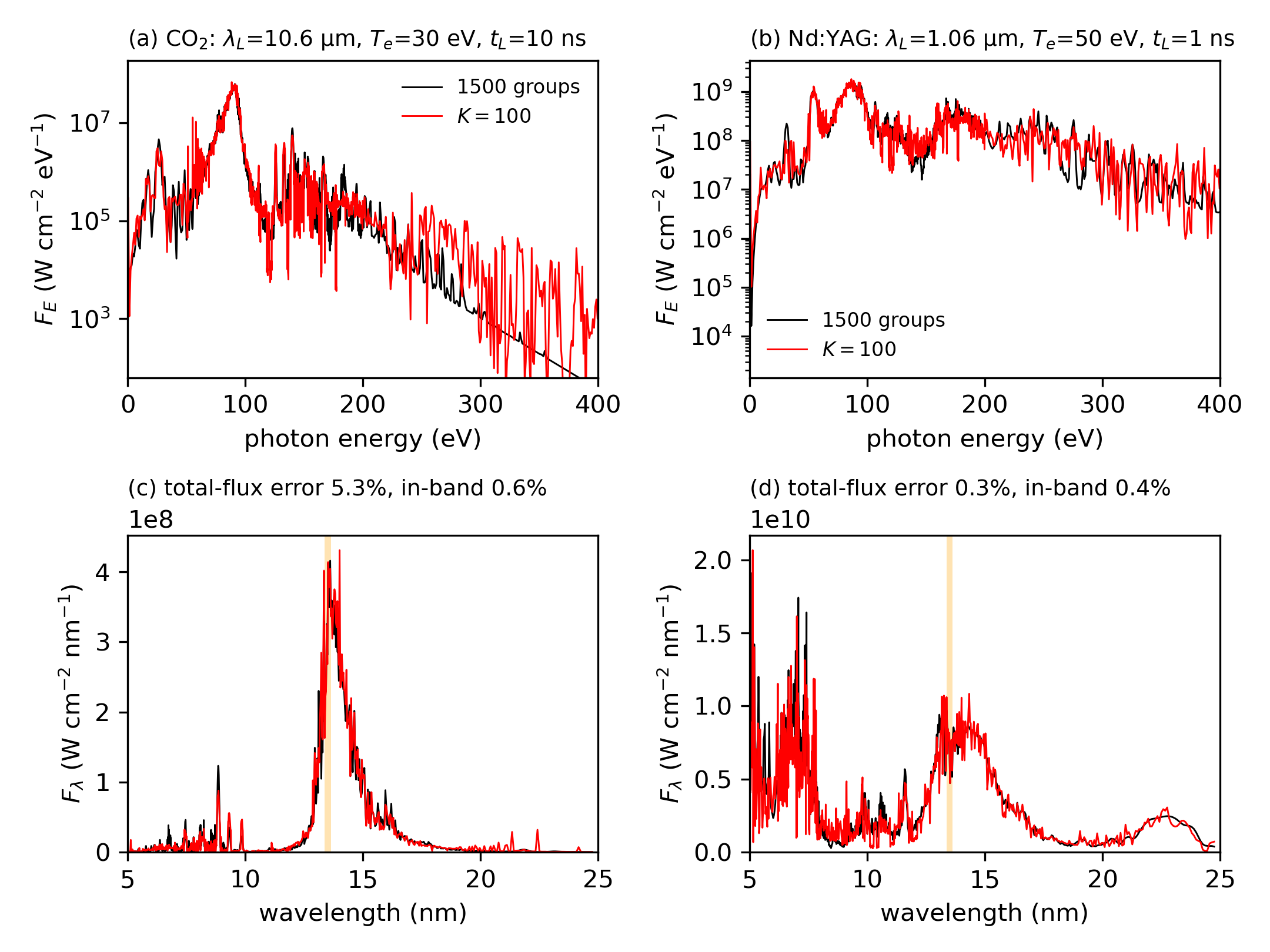}
  \caption{Boundary spectra for (a,c) a CO$_2$-laser condition, $\lamL=10.6\um$, $\Te=30$~eV, $\tL=10$~ns, and (b,d) a Nd:YAG-laser condition, $\lamL=1.06\um$, $\Te=50$~eV, $\tL=1$~ns. Top: as a function of photon energy (logarithmic scale). Bottom: as a function of wavelength (5--25~nm). The shaded band shows 13.5~nm~$\pm1\%$. Black: 1500-group reference. Red: $K=100$. The errors of the total flux and of the in-band flux are 5.3\% and 0.6\% for the CO$_2$ case, and 0.3\% and 0.4\% for the Nd:YAG case.}
  \label{fig:co2}
\end{figure}

\textit{Accuracy and speed.}---Table~\ref{tab:compare} shows the results for the test cases.
With $K=100$, the median error of the total flux is 2.1\% (p90: 8.0\%), the p90 error of the in-band flux is 7.1\%, and the median $\relL$ is 15\%.
With the same number of \emph{contiguous} bands, the p90 error of the total flux is 45\% with the same rank-one shapes and 86\% with flat shapes.
These large errors have a physical reason.
A contiguous band in the UTA region always contains both saturated line cores, where self-absorption keeps $u_g$ close to the source function, and transparent windows, where $u_g$ continues to increase with the plasma thickness.
One amplitude cannot describe both.
The band means are then inconsistent: $\kabs_k$ is determined by the lines, but the harmonic mean $\kDk$ is determined by the windows.
As a result, the band emits like a line and lets the radiation escape like a window, and the flux is too large by up to a factor of two.
When cores and windows are put into different groups, the inconsistency disappears. This is why the non-contiguous groups alone reduce the error by a factor of 5--10.
Non-contiguous groups alone are not enough.
Opacity binning at one state (30~eV, $10^{20}$~cm$^{-3}$) is as poor as contiguous bands (p90 of $\etot$: 82\%), because the temperature changes by a factor of 50 and the density by eight orders of magnitude.
Groups built from the state dependence of $\kappa_g$ are much better (11\%).
With our rank-one shapes, they come close to the proposed model for the total flux (median 2.3\% compared with 2.1\%).
The groups built from the radiation field are better where the line shape matters: the median $\relL$ is 0.15 instead of 0.21, and the p90 in-band error is 7\% instead of 20\%.
The ratio between a saturated core and the windows next to it is set by transport and is not a function of $\kappa_g(\Te,\nion)$.
Only the conventional model has a smaller in-band error (4.5\%).
Its narrow bands in the optically thick UTA peak are almost exact there, but its total flux fails (Supplemental Material).
The computing time is 0.018~s per case, compared with 1.03~s for the reference.
The solver parts alone differ by a factor of about $15\approx G/K$, with the same number of Picard iterations (6--8).
Figure~\ref{fig:co2} shows two conditions that are typical for EUV sources: a plasma produced by a CO$_2$ laser ($\lamL=10.6\um$, $\Te=30$~eV, $\tL=10$~ns) and a plasma produced by a Nd:YAG laser ($\lamL=1.06\um$, $\Te=50$~eV, $\tL=1$~ns). In the second case, the critical density is 100 times higher and the self-absorption of the UTA is much stronger.
The UTA peak and the main lines are reproduced.
The remaining differences are in weak lines and in the high-energy tail, where the flux is orders of magnitude lower than the peak.

\textit{Origin of the remaining error.}---The remaining error comes from the basis, not from the solver.
We projected the reference spectra directly onto the $K=100$ basis. The resulting $\relL$ was the same as that of the reduced transport calculation.
The reason is that transport changes the spectrum \emph{inside} a group. When the optical depth increases along the plasma, the line cores saturate, but the wings and the windows continue to grow.
The shape $w_k$, however, is fixed, and the solver cannot produce a shape that the basis does not contain.
Shapes that depend on the local state $(\Te,\nion)$ do not solve this: even the best local shape reduced the median $\relL$ only from 15\% to about 8\% (Supplemental Material).
The rest of the error is non-local. It is determined by the radiation that comes from upstream.
Integrated quantities, such as the total flux and the in-band flux, are much less affected, because the shape errors mostly cancel inside a group.
A model with $m\ge2$ coupled shapes per group, for example one for the core and one for the wings, is more accurate (Table~\ref{tab:compare}).
However, the flux limiter must then be calculated for each fine bin (cost $\propto G$), and most of the speed-up is lost. We therefore use $m=1$, for which the cost is exactly proportional to $K$.

\textit{Application to a radiation-hydrodynamic code.}---We tested the method in the one-dimensional radiation-hydrodynamic code STAR-1D.\cite{Sunahara2008,Tanaka2026}
This code solves time-dependent multigroup FLD together with the hydrodynamics. The plasma has temperature gradients, and the radiation changes the plasma.
The code uses a table with 151 contiguous groups with flat shapes in this study.
The training spectra were the time-integrated spectra at the plasma boundary of $7000$ runs with this table, chosen at random from the more than $1.47\times10^{5}$ runs of a parameter survey of spherical Sn targets.\cite{Tanaka2026} From these spectra, we built a table with $K=30$ generalized groups.
Only the table reader and the diffusion coefficient of the code were changed (Supplemental Material).
We then repeated 567 runs of the survey with the 30-group table. These runs were excluded from the training. They cover laser wavelengths of 1--11$\um$ and laser intensities of $10^{9}$--$10^{11}$~W/cm$^{2}$, and each is the run with the highest conversion efficiency for its wavelength and intensity.
The conversion efficiency from laser energy to in-band EUV energy (13.5~nm, 2\% bandwidth, $2\pi$~sr) has a maximum of 5.6\% in the survey.\cite{Tanaka2026} 
With the 30-group table, it changed by at most 0.39\% in absolute value (mean $-0.12\%$; Fig.~\ref{fig:star1d_ce}).
A conventional table with 30 contiguous groups and flat shapes, built from the same training data, changed it by a similar amount (at most 0.48\%, mean $+0.16\%$), as expected from the in-band errors in Table~\ref{tab:compare}.
The spectra differ (Fig.~\ref{fig:star1d_spec}): the median $\relL$ of the time-integrated spectrum is 13\% with the generalized groups and 21\% with the conventional groups, and the conventional groups lose 90\% of the emission at 6--13~nm.
The runs were about five times faster (3~min instead of 16~min per run).
The reference here is the 151-group calculation, not a 1500-group calculation.

\textit{Outlook.}---The final model consists of $K$ uncoupled FLD groups with non-contiguous supports and tabulated shape-weighted means.
Nothing in it is specific to one dimension, to a steady state, or to tin.
To use it in a multidimensional radiation-hydrodynamic code, one replaces the opacity table by the $K$-group table and sets the number of groups to $K$.
The time for radiation transport then decreases by a factor $G/K$, and no projection or reconstruction is needed during the simulation.
The training data must represent the target simulation.
The next step is to run a one-dimensional radiation-hydrodynamic code with about $10^{3}$ groups for a few percent of the planned parameter space, to build the Gram matrix from its spectra, and to give the $K$-group table to a two- or three-dimensional code.
The method also answers a general question: which variables of a spectral or kinetic operator can be merged without coupling the merged variables to each other?
Our answer is to merge only the variables that the data show to change together, wherever they are in the spectrum.

\begin{acknowledgments}
The authors thank the technical support staff for their assistance with computer simulations.
This work was supported by the Japan Science and Technology Agency (JST) K Program Grant Number JPMJKP24M1; 
the Joint Usage/Research Center Program of the Institute of Laser Engineering (ILE) at The University of Osaka; the ``Power Laser DX Platform'' as shared research equipment under the Ministry of Education, Culture, Sports, Science and Technology (MEXT) Project for Promoting Public Utilization of Advanced Research Infrastructure (Program for Advanced Research Equipment Platforms, Grant No. JPMXS0450300024); and the Japan Society for the Promotion of Science (JSPS) Core-to-Core Program (Grant No. JPJSCCA20230003).
\end{acknowledgments}

\section*{Author declarations}
\textit{Conflict of interest.}---The authors have no conflicts to disclose.

\textit{Author contributions.}---S.F.: conceptualization (equal), methodology (lead), software (lead), formal analysis (lead), investigation (lead), visualization (lead), validation (equal), funding acquisition (equal), project administration (equal), writing/original draft preparation (lead), writing/review \& editing (lead).
Y.T.: conceptualization (equal), methodology (supporting), validation (supporting), writing/review \& editing (supporting).
N.T.: validation (equal), writing/review \& editing (supporting).
A.S.: data curation (lead; opacity and emissivity table), conceptualization (supporting), validation (supporting), writing/review \& editing (supporting).
Y.Y.: software (supporting), validation (supporting), writing/review \& editing (supporting).
M.F.S.: validation (supporting), writing/review \& editing (supporting).
K.N.: conceptualization (supporting), supervision (equal), writing/review \& editing (supporting).
T.J.: supervision (equal), writing/review \& editing (supporting).
A.Su.: supervision (equal), writing/review \& editing (supporting).
M.Y.: funding acquisition (equal), project administration (equal), writing/review \& editing (supporting).

\textit{Use of artificial intelligence.}---A large language model (Claude, Anthropic) was used, under the direction of the corresponding author, to write the computer code, to help with the mathematical formulation, and to check the English of the manuscript. The authors verified all code, derivations, and numerical results. The corresponding author is fully responsible for the content of this Letter.

\section*{Data availability}
The original tin opacity and emissivity $G$-group and the reduced $K$-group tables cannot be made publicly available.
The Python implementation (group construction, table generation, and $K$-group transport solver) is openly available at \url{https://doi.org/10.5281/zenodo.24163415}.
Because the tin table cannot be distributed, the archive includes a generator of a synthetic 1500-group table for a fictitious element, so that all steps of the method can be run.
The numerical results of this Letter cannot be fully reproduced with the synthetic table.

\bibliographystyle{apsrev4-2}
\bibliography{261114ncg_rad_transport_model}

\clearpage
\onecolumngrid
\setcounter{section}{0}\setcounter{figure}{0}\setcounter{table}{0}\setcounter{equation}{0}
\renewcommand{\thesection}{S\arabic{section}}
\renewcommand{\thetable}{S\arabic{table}}
\renewcommand{\thefigure}{S\arabic{figure}}
\renewcommand{\theequation}{S\arabic{equation}}
\begin{center}
{\large\bfseries Supplemental Material}\\[4pt]
{\bfseries Data-driven non-contiguous photon-energy groups for fast multigroup radiation transport in line-dominated plasmas}
\end{center}

Equation and table numbers without the prefix S refer to the main text.

\section{Reference solver and computing time}
The reference calculation solves Eq.~(1) on $N=500$ cells with $\Delta x=1\um$, using cell-centered finite volumes.
The opacities and emissivities are interpolated from the table with cubic splines in $\log_{10}\Te$ and $\log_{10}\nion$.
The interpolation in $\Te$ is done once, because $\Te$ is constant in the plasma. The interpolation in $\nion$ is done for each cell.
The interpolated values of $\log\kappa_g$ and $\log\varepsilon_g$ are limited to the range of the table for each bin, to avoid overshoot of the splines.
We use a lower limit $\kappa_g\ge10^{-5}$~cm$^{-1}$.
With a smaller limit, the condition number of the tridiagonal systems, about $D_g/(\Delta x^{2}c\kappa_g)$, becomes larger than $10^{12}$. The solution then has round-off noise of about $10^{-4}$, and the Picard iteration does not converge.
Cells with $\nion<10^{15}$~cm$^{-3}$ (the lower limit of the table) are treated as vacuum and are not calculated. In planar geometry, the flux is conserved in vacuum, so the flux at the last calculated cell is the flux at $x_{\mathrm b}$.
The $G$ tridiagonal systems are solved at the same time with the Thomas algorithm.
The flux limiter is treated by Picard iteration. After each linear solve, $R_g$ is calculated on the cell faces from the new $U_g$, and $D_g$ is updated.
The value of $R$ at the boundary is taken from the last interior face. If it is taken from the Marshak boundary value, it depends on itself and the iteration does not work.
If the error of the iteration increases, $D_g$ is relaxed as $D\leftarrow D_{\mathrm{old}}^{1-\omega}D_{\mathrm{new}}^{\omega}$, and $\omega$ is reduced from 1 down to 0.2.
The iteration stops when two changes are below $10^{-5}$: the relative change of $U_g$, weighted by $\kappa U$ (maximum over the cells, normalized for each group), and the relative change of the boundary flux. This needs 6--8 iterations.

On one core (Python/NumPy), one reference case takes 1.03~s (median of the test cases).
About 0.7~s is the interpolation of the table (1500 bins in each cell), and 0.2--0.3~s is the solver with Picard iteration.
The model with $K=100$ takes 0.018~s in total, with the same number of Picard iterations.
Its table interpolation is cheaper by a factor $K/G$, and its solver is the same algorithm for $K$ groups instead of $G$ groups.
The ratio of the total times (58) is therefore larger than the gain of the solver itself, which is about $15\approx G/K$.
The ratio of the total times depends on the cost of the opacity interpolation in each code. The ratio of the solver times does not.

\section{Coupled mode expansions used for comparison}
\textit{POD--Galerkin, projected at each iteration.}
The basis $\phi_k$ is the $\Delta E$-weighted POD of the same training spectra. The modes are the eigenvectors of $\mathsf A$ [Eq.~(5)], divided by $\sqrt{\Delta E_g}$.
At each Picard iteration, the spectrum with $G$ bins is reconstructed in every cell, $R_g$ and $D_g$ are calculated for each bin, and $\mathfrak D$ is projected again. This needs $N\times G\times K^{2}$ operations.
The operators $\mathcal K$ and $\mathcal S$ are projected only once from $\kappa_g$ and $\varepsilon_g$ in the cells.
The $K\times K$ block-tridiagonal system is solved with a banded LAPACK solver with pivoting.
A block Thomas algorithm without pivoting loses 3--4 digits, and the Picard iteration then does not converge. For the same reason, the projection must be done in double precision.
With $K=100$, one case takes 167~s (20--200 iterations, about 2~s each).

\textit{POD--Galerkin, tabulated operators and DEIM limiter.}
The quantities $\mathcal K/\nion$, $\mathcal S/\nion^{2}$, and $\nion\mathfrak D^{0}$ (the diffusion operator without limiter) are projected once on the $(\Te,\nion)$ grid of the table. They are interpolated with local cubic (Catmull--Rom) weights.
The construction of the operators then costs 0.12~s per case and does not depend on $G$.
The limiter factor $f_g=3\lambda(R_g)\in(0,1]$ is expanded in its own POD basis. We used $M=32$ modes; five modes already contain 99.99\% of its energy on the training faces.
The coefficients are fitted by least squares to $P=128$ sample bins, which are selected by pivoted QR (discrete empirical interpolation).
In sample bins where the reconstructed $u_g$ is below $10^{-2}$ of the maximum in the cell, the reconstruction can be close to zero or negative, and $R_g$ becomes very large. In such bins, $f_g$ is replaced by the median of the reliable sample bins on the same face. Without this replacement, the Picard iteration diverges.
The block system is solved with a block Cholesky--Thomas recursion (the Galerkin operator is symmetric and positive definite). This is 3.5 times faster than banded LU.
The cost is still 12~s per case. The block solve takes 0.34~s per iteration.
The slow part is therefore the dense $K\times K$ coupling, not the construction of the operators.

\section{Definition of the contiguous-band models and full results}
The two comparison models use the same solver, the same tables, and the same flux limiter as the proposed model. Only $S_k$ and $w_k$ are different.
\begin{enumerate}\sloppy
  \item \textit{Contiguous bands.} The bins are ordered by photon energy. The cumulative training energy $\sum_{g'\le g}\mathsf A_{g'g'}$ is divided into $K$ equal parts, and band $k$ contains the consecutive bins in the $k$-th part. Every band therefore has the same training energy. This is the best contiguous choice for a given $K$: it gives many narrow bands in the UTA region and a few wide bands elsewhere. 
   \item \textit{Rank-one shapes on contiguous bands.} For each band, $w_k$ is the largest eigenvector of $\mathsf A_{S_kS_k}$ divided by $\sqrt{\Delta E_g}$, in the same way as for the non-contiguous groups. The group means $\kabs_k$, $\ebar_k$, and $\kDk$ are given by Eq.~(4).
  \item \textit{Flat shapes on contiguous bands} (``conventional''). The shape is $w_k(g)=(\sum_{g\in S_k}\Delta E_g)^{-1/2}$. Equation~(4) then gives the $\Delta E$-weighted arithmetic mean of $\kappa_g$ for absorption, the $\Delta E$-weighted mean of $\varepsilon_g$ for emission, and the $\Delta E$-weighted harmonic mean of $\kappa_g$ for diffusion. These are the Planck-type and Rosseland-type band means of a conventional multigroup code with a flat weighting function. A flat weighting function is appropriate when the band is narrow compared with the scale on which $B_E(T)$ changes. This is the case here: for $K=100$, the bands in the UTA region are at most a few eV wide.
\end{enumerate}
The explained energy fraction $\eta$ of each model (Table~I) is the fraction of the training energy that its own basis contains. It is $\sum_k\lambda_{\max}(\mathsf A_{S_kS_k})/\mathrm{tr}\,\mathsf A$ for the rank-one shapes, and $\sum_k v_k^{\mathsf T}\mathsf A_{S_kS_k}v_k/\mathrm{tr}\,\mathsf A$ with the flat $v_k$ for the conventional model.

\begin{figure}[h]
  \centering
  \includegraphics[width=0.8\linewidth]{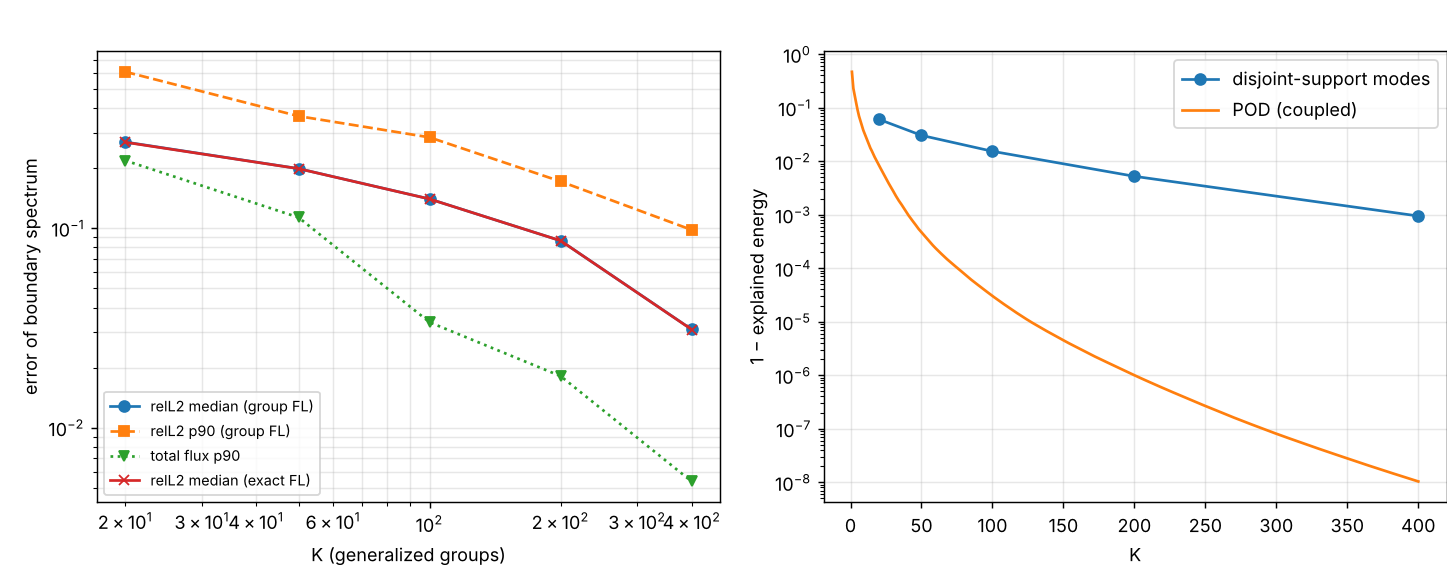}
  \caption{(Left) Errors of the boundary spectrum for the 12 validation cases as a function of $K$: median and 90th percentile of $\relL$, 90th percentile of $\etot$, and median $\relL$ when the flux limiter is calculated for each fine bin instead of each group. (Right) $1-\eta$ for the basis with non-overlapping supports, compared with $1-$(cumulative energy) for the coupled POD basis.}
  \label{fig:scan}
\end{figure}

Table~\ref{tab:full} gives all error values for the models of Table~I, and Table~\ref{tab:kscan} and Fig.~\ref{fig:scan} give the dependence on $K$ for the 12 validation cases.
The row for 400 groups in Table~I was calculated again for the 20 test cases. Its $\etot$ p90 is larger there (3.8\%, compared with 0.5\% for the validation cases), because the test set contains several cases with $\lambda_{\mathrm L}=1\um$ and strong self-absorption. The medians agree.

\begin{table}[h]
  \caption{All error values for the 20 test cases (median and 90th percentile). The two POD--Galerkin rows use the same POD basis with $K=100$. ``Block-diag.\ $K\times m$'' is the block-diagonal model with $K$ generalized groups and $m$ shapes per group. ``conv.'' is the fraction of cases in which the Picard iteration reached its tolerance ($10^{-4}$ for the coupled models, $10^{-3}$ for the block-diagonal model, $10^{-5}$ for the others) within the maximum number of iterations. The errors of the cases that did not converge are not larger than those of the converged cases.}
  \label{tab:full}
  \scriptsize
  \setlength{\tabcolsep}{4pt}
  \begin{ruledtabular}
  \begin{tabular}{lcccccccc}
    Model & $\eta$ & $\relL$ med & $\relL$ p90 & $\etot$ med & $\etot$ p90 & $\eib$ p90 & Time (s) & conv. \\
    \hline
    POD--Galerkin, dense & 0.99997 & 0.011 & 0.073 & 0.001 & 0.006 & 0.033 & 167 & 0.80 \\
    POD--Galerkin, tabulated + DEIM & 0.99997 & 0.036 & 0.080 & 0.006 & 0.014 & 0.048 & 12 & 0.75 \\
    Block-diag., $100\times2$ & 0.997 & 0.071 & 0.216 & 0.011 & 0.056 & 0.020 & 0.50 & 1.00 \\
    Block-diag., $100\times3$ & 0.999 & 0.038 & 0.122 & 0.002 & 0.015 & 0.018 & 0.75 & 1.00 \\
    Block-diag., $50\times4$ & 0.999 & 0.054 & 0.158 & 0.006 & 0.027 & 0.038 & 0.58 & 1.00 \\
    Block-diag., $200\times2$ & 0.9996 & 0.028 & 0.103 & 0.003 & 0.013 & 0.014 & 0.91 & 1.00 \\
    \hline
    Non-contig., rank-one, $K=100$ (this work) & 0.985 & 0.149 & 0.339 & 0.021 & 0.080 & 0.071 & 0.018 & 1.00 \\
    Contiguous, rank-one, $K=100$ & 0.934 & 0.314 & 0.698 & 0.194 & 0.446 & 0.070 & 0.017 & 1.00 \\
    Contiguous, flat, $K=100$ & 0.649 & 0.552 & 0.830 & 0.137 & 0.864 & 0.045 & 0.016 & 1.00 \\
    Non-contig., rank-one, $K=200$ & 0.995 & 0.097 & 0.244 & 0.016 & 0.067 & 0.034 & 0.029 & 1.00 \\
    Non-contig., rank-one, $K=400$ & 0.999 & 0.041 & 0.144 & 0.004 & 0.038 & 0.007 & 0.051 & 1.00 \\
  \end{tabular}
  \end{ruledtabular}
\end{table}

\begin{table}[h]
  \caption{Dependence on the number of generalized groups $K$ (12 validation cases, flux limiter for each group). In the last two columns (``b-b-b'', bin by bin), the flux limiter is calculated for each fine bin with the full $\kappa_g$ instead of the scalar $R_k$. ``POD'' is the cumulative energy of the coupled POD basis with the same number of modes.}
  \label{tab:kscan}
  \scriptsize
  \setlength{\tabcolsep}{4pt}
  \begin{ruledtabular}
  \begin{tabular}{rccccccccc}
    $K$ & $\eta$ & POD & $\relL$ med & $\relL$ p90 & $\etot$ p90 & $\eib$ p90 & Time (s) & $\relL$ med (b-b-b) & $\etot$ p90 (b-b-b) \\
    \hline
    20  & 0.9394 & 0.9915   & 0.270 & 0.609 & 0.218 & 0.387 & 0.014 & 0.270 & 0.206 \\
    50  & 0.9697 & 0.9995   & 0.199 & 0.365 & 0.113 & 0.272 & 0.016 & 0.199 & 0.073 \\
    100 & 0.9847 & 0.99997  & 0.140 & 0.285 & 0.034 & 0.058 & 0.019 & 0.140 & 0.019 \\
    200 & 0.9948 & 0.999999 & 0.087 & 0.172 & 0.018 & 0.025 & 0.029 & 0.087 & 0.018 \\
    400 & 0.9991 & 1.000000 & 0.031 & 0.098 & 0.005 & 0.017 & 0.050 & 0.031 & 0.011 \\
  \end{tabular}
  \end{ruledtabular}
\end{table}

\section{Non-contiguous groups built from the opacity}\label{sec:sopac}
These partitions use only the opacity table. No transport solution is used to define $S_k$.
All of them use the training energy $\mathsf A_{gg}$ as the weight of bin $g$ in the clustering, with the same $k$-means++ initialization and the same random seed as the proposed model. The $K$ groups are therefore distributed over the spectrum in a comparable way.
\begin{enumerate}
  \item \textit{Opacity binning} (``$\kappa$ value''). Weighted one-dimensional $k$-means of $\log_{10}\kappa_g$ at one reference state, $\Te=30$~eV and $\nion=10^{20}$~cm$^{-3}$, which is typical for EUV emission. This is the picket-fence or opacity-distribution-function construction:\cite{Nordlund1982,Mihalas1984,Auer2000,Wray2006} bins with a similar opacity are put into one group, wherever they are in photon energy.
  \item \textit{State dependence of the opacity} (``$\kappa$ dependence''). For each bin, we form the vector $\log_{10}\kappa_g(\Te,\nion)$ over the whole table ($44\times81$ states) and subtract its mean over the states. The bins are clustered by weighted kernel $k$-means with the cosine similarity of these vectors as kernel. This is the idea of the multigroup full-spectrum $k$-distribution method.\cite{Zhang2003}
  \item \textit{State dependence and value} (``$\kappa$ dep.+value''; the rows ``$\kappa$ state dependence'' in Table~I). As item 2, but the mean of $\log_{10}\kappa_g$ over the states is added to the vector, with the same weight as the dependence part.
  \item \textit{State dependence of opacity and emissivity} (``$\kappa,\varepsilon$ dep.+value''). As item 3, with the state dependence of $\log_{10}\varepsilon_g$ added.
\end{enumerate}
For each partition, the shape inside a group is either flat, as in all opacity-binning and $k$-distribution methods, or the rank-one shape of Eq.~(3) calculated from the training Gram matrix. The second choice is a hybrid: it uses transport data for the shape but not for the groups.

\begin{figure}[h]
  \centering
  \includegraphics[width=0.6\linewidth]{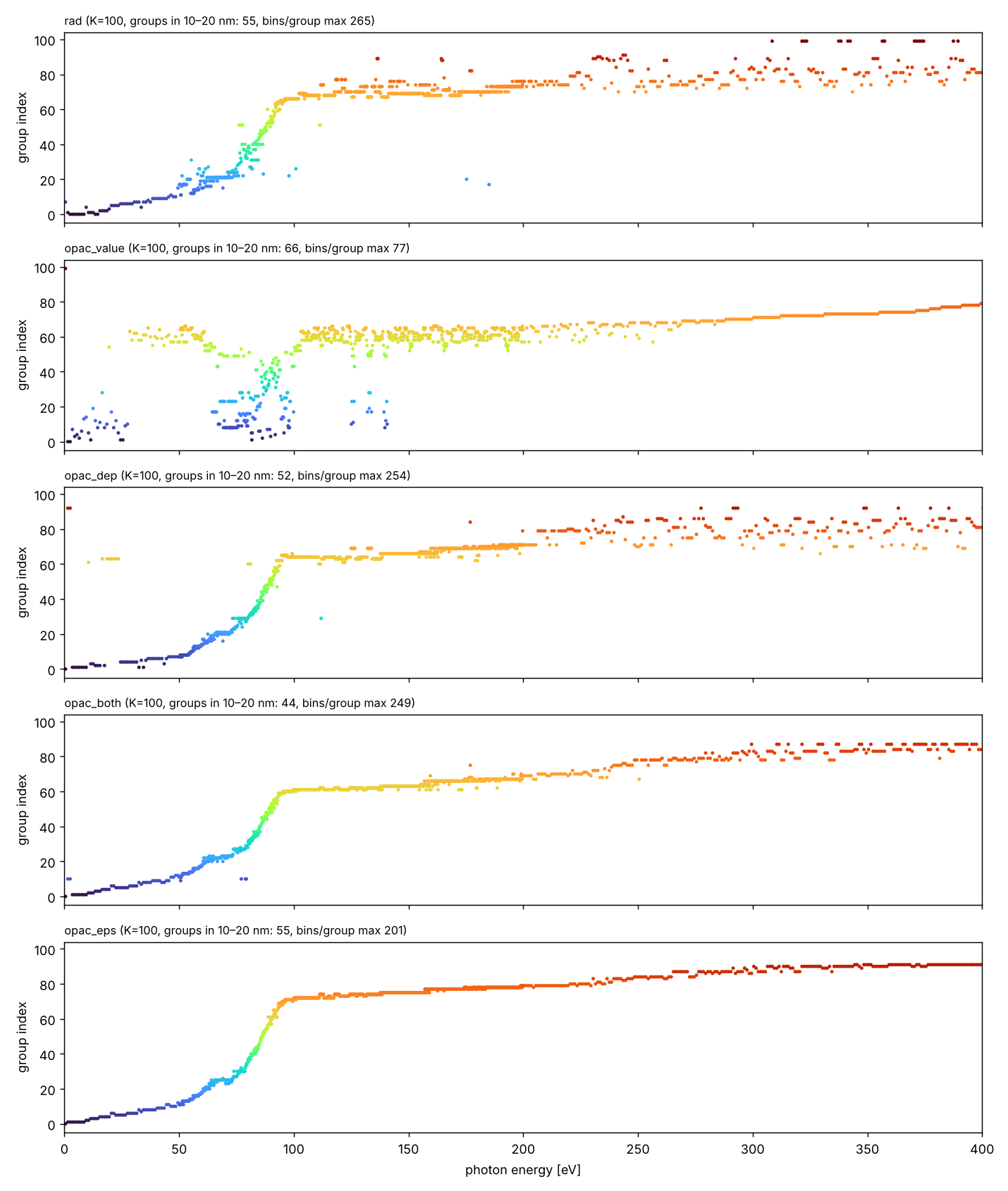}
  \caption{Partitions of the 1500 fine bins into $K=100$ groups (group index, ordered by mean photon energy, as a function of photon energy). From top to bottom: radiation field (this work), $\kappa$ value (opacity binning), $\kappa$ dependence, $\kappa$ dependence and value, and $\kappa,\varepsilon$ dependence and value.}
  \label{fig:parts}
\end{figure}

Table~\ref{tab:base} gives the results, and Fig.~\ref{fig:parts} compares the partitions.
Four points follow.
(i)~Non-contiguous groups alone are not enough. Opacity binning at one state is as poor as contiguous bands, and its in-band error is the largest of all models. A set of bins with the same $\kappa$ at 30~eV and $10^{20}$~cm$^{-3}$ does not have the same $\kappa$ at other states.
(ii)~Groups built from the state dependence of $\kappa_g$ are much better. With the value included, the p90 of $\etot$ is 11\% with flat shapes.
(iii)~The rank-one shape improves every partition. With it, the groups built from the opacity come close to the proposed model for the total flux (median 2.3\% compared with 2.1\%).
(iv)~The groups built from the radiation field are better for the line shape and for the narrow band: the median $\relL$ is 0.15 instead of 0.21, and the p90 of $\eib$ is 7\% instead of 20\%.

\begin{table}[h]
  \caption{Uncoupled models with $K=100$ that differ only in the construction of the groups and in the shape (20 test cases; median and 90th percentile). ``Band $\relL$'' is the median $\relL$ within 10--20~nm. $N_{10\text{--}20}$ is the number of groups in 10--20~nm.}
  \label{tab:base}
  \scriptsize
  \setlength{\tabcolsep}{4pt}
  \begin{ruledtabular}
  \begin{tabular}{llcccccccc}
    Groups from & Shape & $\eta$ & $\relL$ med & $\relL$ p90 & $\etot$ med & $\etot$ p90 & $\eib$ p90 & Band $\relL$ & $N_{10\text{--}20}$ \\
    \hline
    Radiation field (this work) & rank-one & 0.985 & 0.149 & 0.339 & 0.021 & 0.080 & 0.071 & 0.171 & 55 \\
    Radiation field & flat & 0.919 & 0.343 & 0.555 & 0.049 & 0.145 & 0.082 & 0.241 & 55 \\
    $\kappa$ value (opacity binning) & rank-one & 0.895 & 0.436 & 0.742 & 0.206 & 0.441 & 1.93 & 0.417 & 66 \\
    $\kappa$ value (opacity binning) & flat & 0.410 & 0.769 & 1.04 & 0.255 & 0.822 & 2.11 & 0.657 & 66 \\
    $\kappa$ dependence & rank-one & 0.968 & 0.234 & 0.398 & 0.032 & 0.168 & 0.216 & 0.240 & 52 \\
    $\kappa$ dependence & flat & 0.793 & 0.448 & 0.719 & 0.066 & 0.218 & 0.286 & 0.329 & 52 \\
    $\kappa$ dep.+value & rank-one & 0.972 & 0.212 & 0.348 & 0.023 & 0.086 & 0.197 & 0.213 & 44 \\
    $\kappa$ dep.+value & flat & 0.897 & 0.367 & 0.453 & 0.033 & 0.111 & 0.199 & 0.362 & 44 \\
    $\kappa,\varepsilon$ dep.+value & rank-one & 0.971 & 0.236 & 0.373 & 0.025 & 0.174 & 0.191 & 0.211 & 55 \\
    $\kappa,\varepsilon$ dep.+value & flat & 0.878 & 0.321 & 0.580 & 0.049 & 0.245 & 0.266 & 0.273 & 55 \\
    Contiguous & rank-one & 0.934 & 0.314 & 0.698 & 0.194 & 0.446 & 0.070 & -- & -- \\
    Contiguous (conventional) & flat & 0.649 & 0.552 & 0.830 & 0.137 & 0.864 & 0.045 & -- & -- \\
  \end{tabular}
  \end{ruledtabular}
\end{table}

\section{In-band error of the conventional model}
In Table~I, the in-band error is smallest for the conventional model (4.5\%). The reason is as follows.
The contiguous bands have equal training energy, so 26 of the 100 bands are in the UTA peak (85--95~eV). The in-band range (nine fine bins) is covered by five bands of one to three bins, and these bands contain no bin outside this range.
The peak is optically thick, and self-absorption smooths the line structure there. A flat shape in such a narrow band is therefore almost exact.
The cost appears elsewhere: only one band covers 95--130~eV, and the total flux fails.
In the proposed model, four of the eight groups in the in-band range also contain bins outside it (at 89.9 and 92.9--95.3~eV). Their fixed shapes cannot follow the different saturation of the core and of the wing, and this gives an in-band error of a few percent. This error decreases to 0.7\% for $K=400$.

\section{Why contiguous bands fail: a model with two bins}
We consider a band with two bins of equal width: a saturated line core (bin 1, opacity $\kappa_1$) and a transparent window (bin 2, opacity $\kappa_2\ll\kappa_1$).
In the reference solution, self-absorption keeps $u_1$ close to the source function $S_1=\varepsilon_1/(c\kappa_1)$. In contrast, $u_2$ increases linearly with the thickness of the emitting plasma until the window becomes optically thick.
One amplitude fixes the ratio $u_2/u_1$.
With flat weights, the band absorption $\kabs\approx\kappa_1/2$ is determined by the line. The diffusion coefficient, proportional to $1/\kDk\approx1/(2\kappa_2)$, is determined by the window.
The band therefore emits with the emissivity of the line, but the radiation escapes with the transparency of the window.
In the limit, the flux can be larger than the reference by a factor of the order of $\kappa_1/\kappa_2$. This explains why the p90 error of the total flux of the conventional model is 86\%.
A rank-one shape gives more weight to the bin that had more energy in the training data. This reduces the inconsistency (45\%), but it cannot remove it, because the ratio $u_2/u_1$ changes with the optical depth.
If the two bins are put into different groups, the problem disappears.

\section{Intra-group shapes that depend on the state}
We tested whether the fixed shape limits the accuracy. Two variants were calculated with the same partition ($K=100$).
\begin{enumerate}
  \item \textit{Regression of the logarithm of the shape.} For each bin, the intra-group shape was normalized to the maximum in its group for each sample, with a lower limit of $10^{-5}$. Its logarithm, $\log_{10}w_k(g;\Te,\nion)$, was fitted by ridge regression to the 56\,540 training samples. The fit is a linear combination of nine functions of the state, $\{1,\ \ell_T,\ \ell_T^{2},\ \ell_T^{3},\ \ell_n,\ \ell_n^{2},\ \ell_T\ell_n,\ 1/\Te,\ \ell_n/\Te\}$, with $\ell_T=\log_{10}\Te$ and $\ell_n=\log_{10}\nion$. The terms with $1/\Te$ represent Boltzmann factors. The $K$-group table was then calculated again with the state-dependent shape at every point of the table. The fit explains 81\% of the variance of $\log w_k$ in the training set. However, the transport error \emph{increases}: the median $\relL$ changes from 15\% to 24\%, and the p90 of $\etot$ from 8\% to 34\%. The reason is that a regression in logarithmic space gives a geometric mean of the shapes. In the groups with large energy, it gives too little weight to the strong bins and too much weight to the weak bins. The rank-one shape, in contrast, is the best shape in the energy norm.
  \item \textit{Best local shape (upper limit).} To estimate the best result that any function of the local state can give, we projected the reference boundary spectra of the test cases onto shapes that were built only from training samples with the same $\Te$ and with a density within $\pm0.5$ decades of the boundary density. The median projection error decreased from 15\% (global shapes) to about 8\%. Several cases with $\lambda_{\mathrm L}=1\um$ stayed near 30\%.
\end{enumerate}
The remaining 8\% is the non-local part of the change of the intra-group shape.
Inside a group, the ratio between the bins at a given point depends on the radiation that arrives from upstream, that is, on how strongly the cores are saturated. This is not a function of the local temperature and density.
A state-dependent shape also needs state-dependent weights $w_k(g;\Te,\nion)$ for the reconstruction, in addition to the $K$-group table. For these reasons, we did not use either variant.

\section{Block-diagonal model: $m$ shapes per group}
In the block-diagonal model, each generalized group $S_k$ has $m_k$ orthonormal shapes with independent amplitudes. The shapes are the first $m_k$ eigenvectors of $\mathsf A_{S_kS_k}$.
The reduced operators are then $m_k\times m_k$ blocks, which are tabulated on the $(\Te,\nion)$ grid.
When the unknowns are ordered as (group, cell, shape), the linear system is a banded matrix with bandwidth $2m-1$.
The total number of shapes, $\sum_k m_k=Km$, is distributed among the groups. Groups with a larger second eigenvalue get more shapes, and each group gets at most $m+1$.
One scalar flux limiter per group cannot be used for $m\ge2$.
If one $\lambda(R_k)$, calculated from the amplitude of the first shape, is applied to the whole $m\times m$ block, the limiter is too strong for the core and too weak for the wings. In the test cases, the median $\relL$ for $100\times2$ then increased from 3.5\% to 33\%, and the iteration did not converge. Without any limiter, the transport error is already equal to the projection error of the basis.
The limiter is therefore calculated for each fine bin from the reconstructed $u_g$ and is projected onto the blocks at every Picard iteration. The cost is proportional to $N\,G\,m^{2}$.
For stability, we used a Picard tolerance of $10^{-3}$ (4--5 iterations) and a lower limit of $10^{-3}$ for the reconstructed $u_g$.
The results for the 20 test cases are given in Table~\ref{tab:full}.
The accuracy of $100\times3$ is equal to that of a coupled POD basis with about 50--60 modes.
In one dimension, however, the limiter for each fine bin keeps the computing time within a factor of 1.4--2 of the reference.
In two or three dimensions, the elliptic solves dominate the cost, and this ratio will be better. The block-diagonal model may be useful there.

\section{Example spectra}
\begin{figure}[h]
  \centering
  \includegraphics[width=0.9\linewidth]{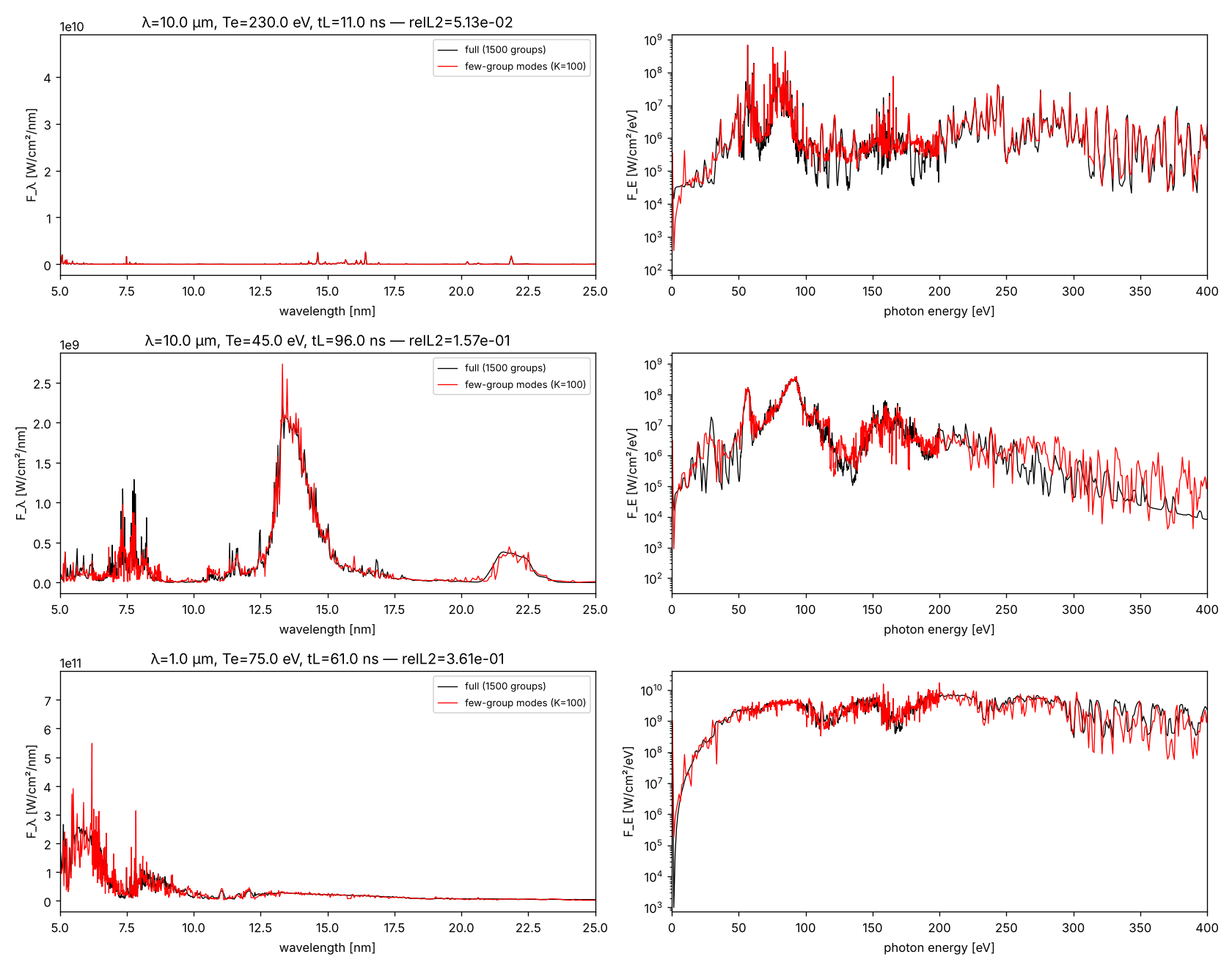}
  \caption{Test cases with the smallest (top), median (middle), and largest (bottom) $\relL$. Left: 5--25~nm. Right: 0--400~eV, logarithmic scale. Black: 1500-group reference. Red: $K=100$.}
  \label{fig:examples}
\end{figure}

Figure~\ref{fig:examples} shows the test cases with the smallest, median, and largest $\relL$ for $K=100$.

\section{Weighting of a band}
We tested two ways to give priority to the 10--20~nm band in the partition for $K=100$. In the first (soft), the inner product is multiplied by a weight larger than 1 inside the band. In the second (hard), a fixed number of groups is reserved for the bins inside the band.
The median $\relL$ inside the band improved only from 17\% to 15\% with a soft weight of 10, and to 11--12\% with 70--85 groups reserved for the band.
At the same time, the $\relL$ of the whole spectrum increased from 15\% to 17--19\%, and the p90 error of the 2\% bandwidth increased from 7\% to 27\%. Soft weights of 100 or more made the whole spectrum much worse.
The reason is that 55 of the 100 groups are already inside the band without any weighting, because the band contains 62\% of the training energy.
It is more effective to increase $K$. With $K=200$ (150 groups inside the band, hard assignment), the $\relL$ inside the band is 3.5\%, the p90 error of the flux integrated over the band is 1.0\%, and the p90 error of the 2\% bandwidth is 0.7\%, at 0.026~s per case.

\section{Application to the radiation-hydrodynamic code STAR-1D}
STAR-1D is a one-dimensional radiation-hydrodynamic code.\cite{Sunahara2008,Tanaka2026} The runs used here are in spherical geometry.
Its radiation transport is time-implicit multigroup FLD. The opacity and emissivity tables are interpolated in $(\ln\nion,\ln\Te)$.
The standard table for Sn has 151 contiguous groups, and the code treats each group with a flat shape.

\textit{Table.} The code stores a spectral energy density for each group. For a generalized group, we use $\tilde u_k=a_kW_k/\Delta S_k$, with $W_k=\sum_{g\in S_k}w_k\Delta E_g$ and $\Delta S_k=\sum_{g\in S_k}\Delta E_g$. This is the mean spectral density over the support of the group.
The 30-group table then contains $\kabs_k$, the emissivity $\ebar_kW_k/\Delta S_k$, the width $\Delta S_k$, and $\kDk$ [Eq.~(4)].
With these definitions, the sums over groups in the code (radiation energy, exchange with the matter, boundary flux) keep their form, and the energy is conserved.
For flat shapes, the table is equal to a usual multigroup table.

\textit{Changes of the code.} Three parts were changed. (i)~The table reader also reads $\kDk$ and a $151\times30$ matrix that reconstructs the 151-group spectrum from $\tilde u_k$. (ii)~The diffusion coefficient uses $\kDk$ instead of the absorption opacity. (iii)~The output spectrum and the in-band energy (13.5~nm, 2\% bandwidth) are calculated from the reconstructed spectrum.
The hydrodynamics, the laser absorption, and the flux limiter were not changed.

\textit{Training.} The parameter survey of Ref.~\cite{Tanaka2026} contains more than $1.47\times10^{5}$ STAR-1D runs with the 151-group table. The target is a Sn sphere with an initial density of 0.04~g/cm$^{3}$. The laser wavelength (1--11$\um$), the laser intensity ($10^{9}$--$10^{11}$~W/cm$^{2}$), the pulse width (0.5--30~ns), the pulse shape, and the target radius (5--1000$\um$) were changed. We chose 7000 of these runs (5\% of the total) at random, after the 567 runs used for the test below had been removed. The training data are the time-integrated spectra at the plasma boundary of these runs, one spectrum for each run. Spectra inside the plasma were not used, in contrast to the main text.
The reason is that the runs of the survey have very different pulse widths, target radii, and laser intensities, so the scale length and the optical depth of the emitting plasma change strongly from run to run.
The boundary spectrum of a thin plasma shows weak self-absorption, and that of a thick plasma shows saturated line cores.
The set of boundary spectra therefore contains the same sequence, from thin to thick, that the radiation field goes through inside one plasma.
In this sense, the change of the scale length between the runs plays a role similar to the sampling at different depths in the main text. The generalized groups and the shapes were then determined from the Gram matrix in the same way as in the main text. 

\begin{figure}[h]
  \centering
  \includegraphics[width=0.245\linewidth]{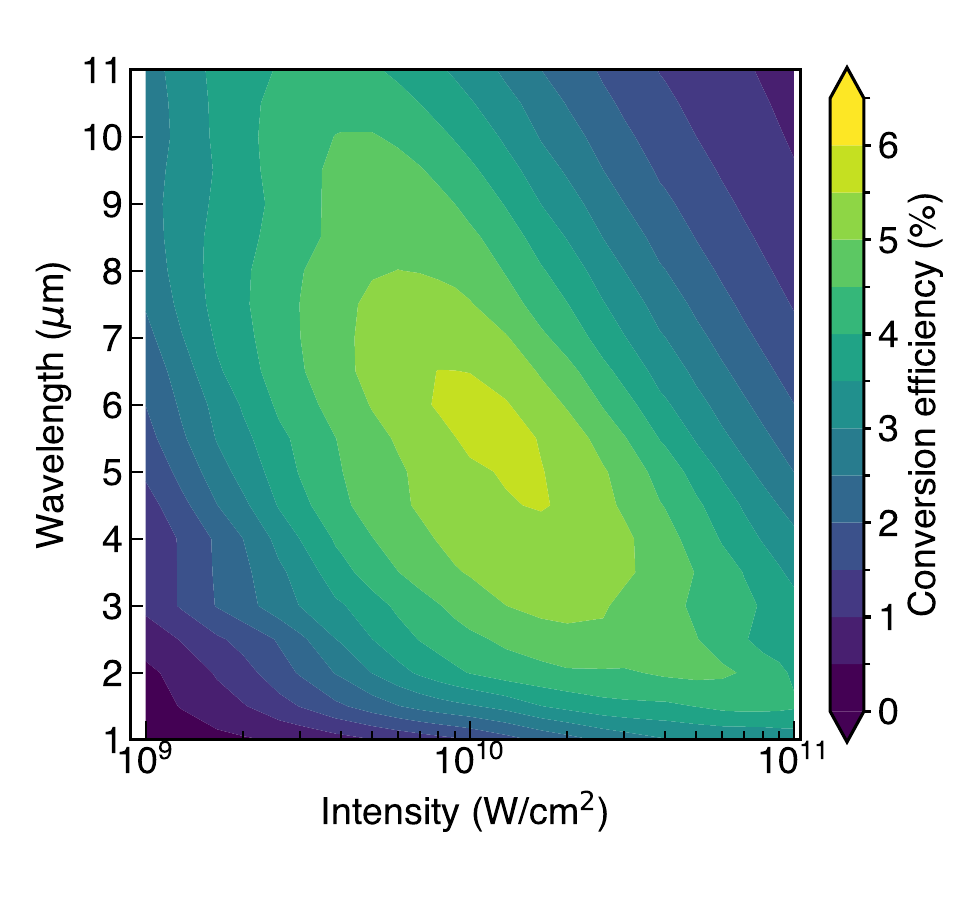}\hfill
  \includegraphics[width=0.245\linewidth]{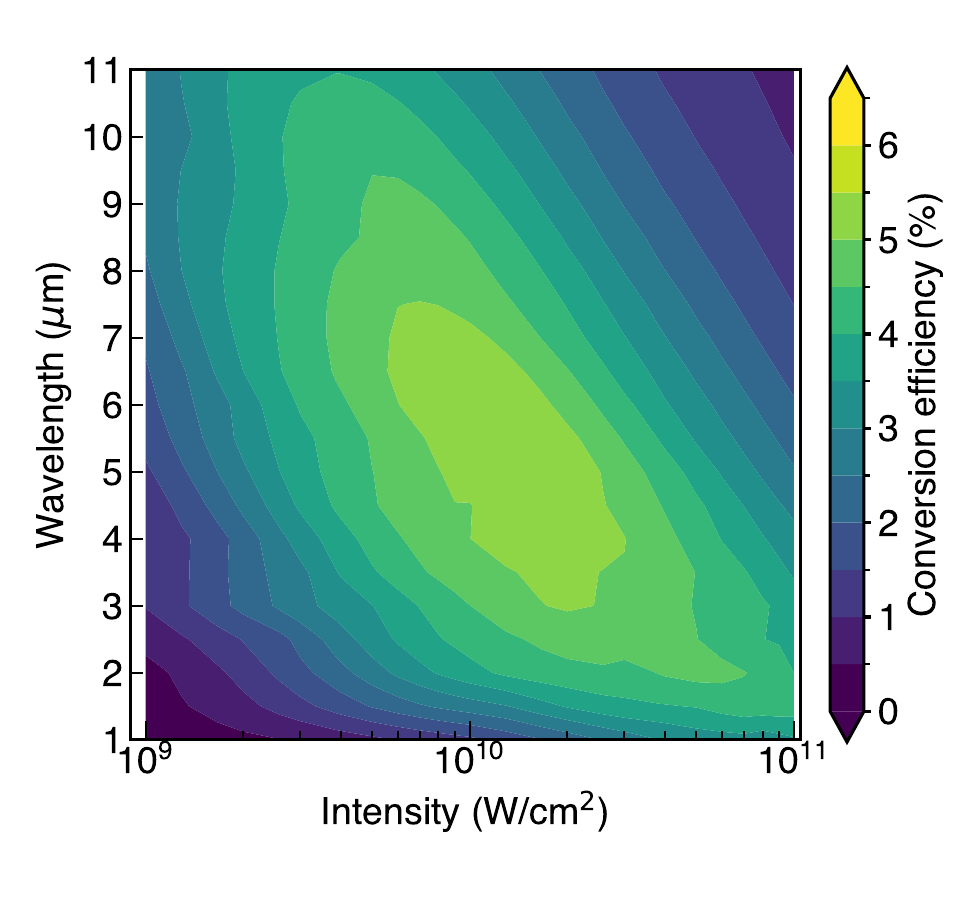}\hfill
  \includegraphics[width=0.245\linewidth]{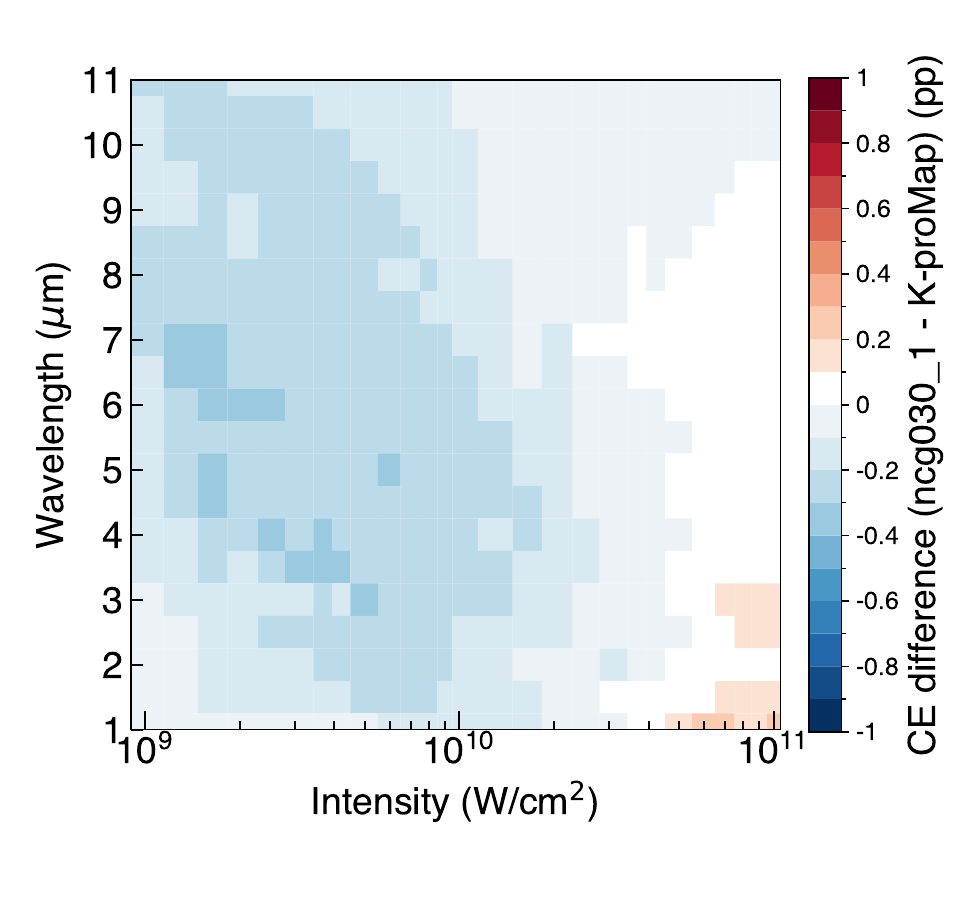}\hfill
  \includegraphics[width=0.245\linewidth]{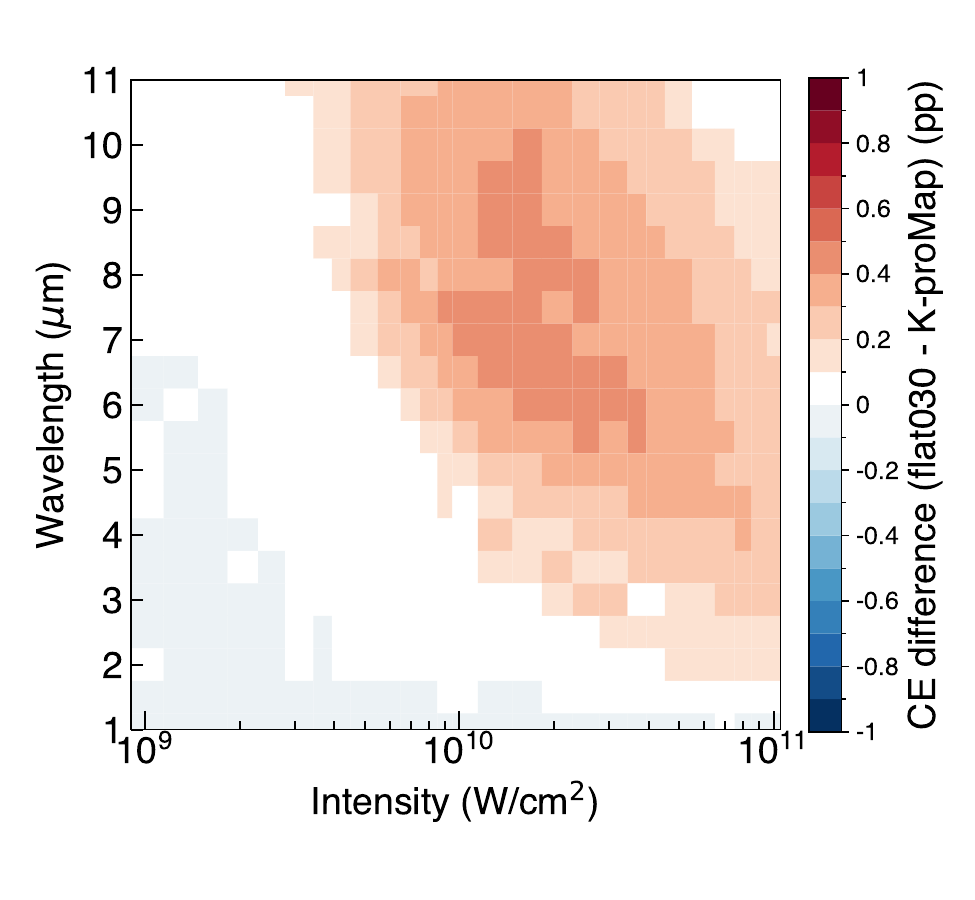}
  \caption{Conversion efficiency calculated with STAR-1D as a function of the laser intensity and the laser wavelength. From left to right: original table with 151 contiguous groups; table with 30 generalized groups; absolute difference between the 30 generalized groups and the 151 groups, in \% (``pp'' on the color bar); the same for 30 conventional groups (contiguous, flat).}
  \label{fig:star1d_ce}
\end{figure}

\begin{figure}[h]
  \centering
  \includegraphics[width=0.75\linewidth]{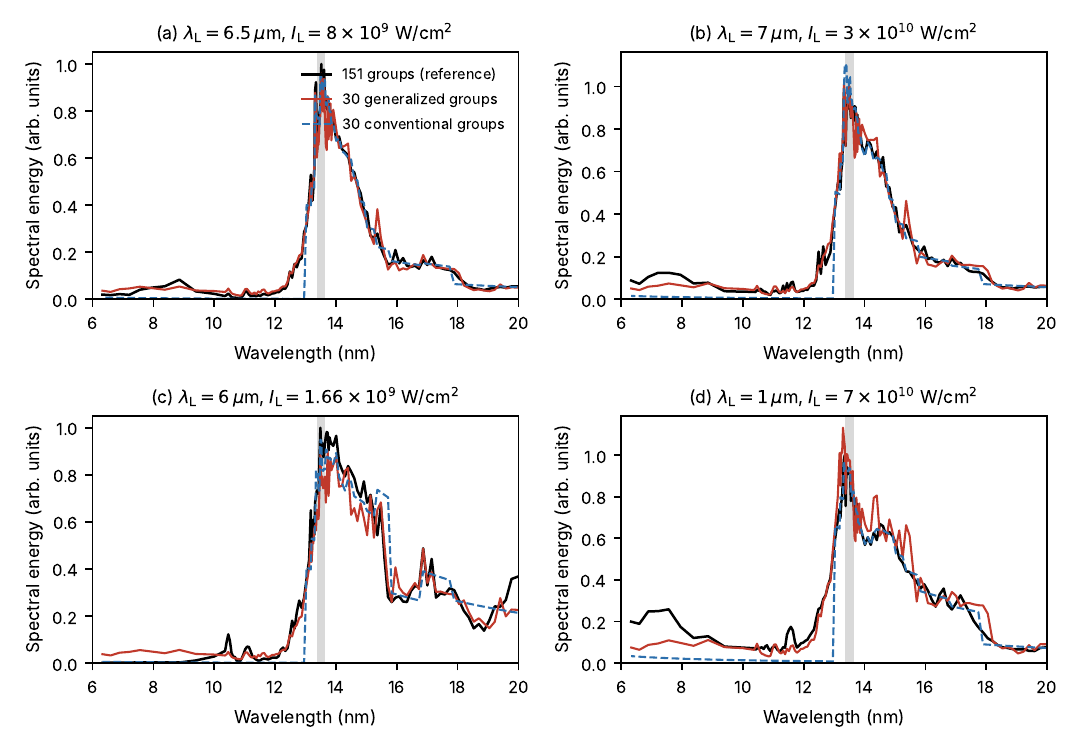}
  \caption{Time-integrated spectra calculated with STAR-1D for four of the 567 runs. Black: original table with 151 contiguous groups. Red: table with 30 generalized groups. Blue dashed: table with 30 conventional groups (contiguous, flat). The spectra with 30 groups are reconstructed on the 151-group mesh. The shaded band shows 13.5~nm~$\pm1\%$. Each panel is normalized to the maximum of the reference.}
  \label{fig:star1d_spec}
\end{figure}

\textit{Results.} The survey has 256 runs for each laser wavelength and each laser intensity. From these, we took the run with the highest conversion efficiency. This gives 567 runs. We repeated these runs with the 30-group table. None of them is in the training set.
Figure~\ref{fig:star1d_ce} shows the conversion efficiency (CE) from laser energy to EUV energy emitted into $2\pi$~sr within a 2\% bandwidth at 13.5~nm. Its maximum with the 151-group table is 5.63\% at a laser wavelength of 5.5$\um$.\cite{Tanaka2026}
The two maps have the same shape, and the maximum is in the same region of wavelength and intensity.
Table~\ref{tab:star1d} gives the statistics of the difference. All differences of the CE and of the radiated fraction in this section are absolute differences, given in \%.
With the 30 generalized groups, the mean difference is $-0.12\%$, and the largest difference is $-0.39\%$ (at $1.66\times10^{9}$~W/cm$^{2}$ and 6$\um$). The CE is slightly lower over most of the map and slightly higher at the highest intensities.
The relative difference of the CE has a median of 3.7\% and a 90th percentile of 9.1\%.
Figure~\ref{fig:star1d_spec} compares time-integrated spectra for four runs.
The UTA peak near 13.5~nm and its long-wavelength side are reproduced. The weak emission at 6--13~nm is reproduced less well; the largest differences are at 6--9~nm for a laser wavelength of 1$\um$.
For all 567 runs, the $\relL$ of the time-integrated spectrum with respect to the 151-group calculation has a median of 13\% and a 90th percentile of 17\%. The spectrally integrated energy agrees within 1.5\% (median).

\textit{Checks.} Two checks were made (Table~\ref{tab:star1d}).
First, the 151-group table was written in the new table format and was used with the modified code. The CE agrees with that of the original code within 0.06\%. This shows that the changes of the code do not change the result.
Second, the 30 generalized groups were also built from all $145\,152$ runs of the survey instead of 7000 runs. The differences of the CE are almost the same (mean $-0.12\%$, largest $-0.35\%$). A training set of 5\% of the runs is therefore sufficient.

\textit{Comparison with 30 conventional groups.} We built a conventional table with 30 contiguous groups and flat shapes from the same 7000 training spectra (contiguous bands with equal training energy, as in Sec.~S3) and repeated the 567 runs.
The difference of the CE has the opposite sign and a similar size: the mean is $+0.16\%$ and the largest difference is $+0.48\%$ (at $1.66\times10^{10}$~W/cm$^{2}$ and 6.5$\um$; Fig.~\ref{fig:star1d_ce}, right).
For the CE, the two 30-group tables are therefore similar.
This agrees with the result of the main text: the contiguous bands of equal training energy are narrow in the UTA peak, and the in-band energy is accurate (Table~I).
The relative difference of the CE has a median of 4.0\% and a 90th percentile of 13.5\% (3.7\% and 9.1\% with the generalized groups).
The difference between the two tables is in the spectrum (Fig.~\ref{fig:star1d_spec} and Table~\ref{tab:star1d}).
With the conventional 30 groups, the spectrum is a step function, and its $\relL$ is larger (median 21\%, 90th percentile 30\%).
The emission at 6--13~nm, which contains 15\% of the energy at 6--20~nm (median), is lower by 90\% in almost all runs. One wide band covers this range and the higher photon energies, and a flat shape cannot tell where in the band the energy is.
With the 30 generalized groups, the median error of the energy at 6--13~nm is 26\%.
The radiated fraction of the laser energy is always too large with the conventional groups (mean $+0.9\%$, up to $+9.8\%$). With the generalized groups, it has no bias (mean $-0.02\%$, between $-4.2\%$ and $+1.5\%$).
We note that the reference of this test has only 151 groups and does not resolve the lines. A larger difference between the two tables is expected when the reference resolves the lines, as in the main text.

\textit{Computing time.} The runs with both 30-group tables took about 3~min per run, compared with about 16~min per run with 151 groups (Table~\ref{tab:star1d}).
The number of groups is smaller by a factor of five, and the run time by a factor of 5.3. The radiation transport therefore takes most of the run time of this code.

\begin{table}[h]
  \caption{STAR-1D results for the 567 test runs, with respect to the original 151-group calculation. $\Delta$CE is the absolute difference of the conversion efficiency in \%. $\relL$ is the weighted relative $L^{2}$ error of the time-integrated spectrum. $\delta E_{6\text{--}13}$ is the median relative error of the energy emitted at 6--13~nm. $\Delta\eta_{\mathrm{rad}}$ is the mean absolute difference of the radiated fraction of the laser energy in \%. ``Time'' is the mean run time (min:s). ``Training'' is the number of training runs. ``30 conventional'' means 30 contiguous groups with flat shapes. ``151 (check)'' is the 151-group table in the new table format. $|\Delta\mathrm{CE}|/\mathrm{CE}$ p90 is the 90th percentile of the relative difference.}
  \label{tab:star1d}
  \scriptsize
  \setlength{\tabcolsep}{4pt}
  \begin{ruledtabular}
  \begin{tabular}{llccccccccc}
    Groups & Training & Time & $\Delta$CE mean & $\Delta$CE min & $\Delta$CE max & $|\Delta\mathrm{CE}|/\mathrm{CE}$ p90 & $\relL$ med & $\relL$ p90 & $\delta E_{6\text{--}13}$ & $\Delta\eta_{\mathrm{rad}}$ \\
    \hline
    30 generalized & 7000 & 3:08 & $-0.12$ & $-0.39$ & $+0.26$ & 0.091 & 0.126 & 0.166 & $0.26$ & $-0.02$ \\
    30 generalized & $145\,152$ & 2:51 & $-0.12$ & $-0.35$ & $+0.26$ & 0.083 & 0.124 & 0.165 & $0.26$ & $+0.00$ \\
    30 conventional & 7000 & 2:50 & $+0.16$ & $-0.09$ & $+0.48$ & 0.135 & 0.211 & 0.295 & $0.90$ & $+0.94$ \\
    151 (check) & -- & 16:27 & $-0.00$ & $-0.06$ & $+0.04$ & 0.006 & 0.000 & 0.011 & $0.00$ & $-0.00$ \\
  \end{tabular}
  \end{ruledtabular}
\end{table}

In summary, this test shows that the method works in a time-dependent code with temperature gradients and with feedback of the radiation on the plasma. The advantage over conventional groups is small for the CE and clear for the spectrum. The test does not give the error with respect to a calculation that resolves the lines.

\end{document}